\documentclass[aps,prl,twocolumn,showpacs,superscriptaddress]{revtex4}
\usepackage{graphicx}

\begin{document}

\title{Deformation Energy Minima at Finite Mass Asymmetry}

\author{D. N. Poenaru}
\email[]{poenaru@ifin.nipne.ro}
\affiliation{
National Institute of Physics and Nuclear
Engineering, RO-76900 Bucharest, Romania}
\affiliation{Institut f\" ur Theoretische Physik der J. W. Goethe 
Universit{\" a}t, D-60054 Frankfurt am  Main, Germany}

\author{W. Greiner}
\affiliation{Institut f\" ur Theoretische Physik der  J. W. Goethe 
Universit{\" a}t, D-60054 Frankfurt am  Main, Germany}

\date{\today}

\begin{abstract}
A very general saddle point nuclear shape may be found as a solution of an
integro-differential equation without giving {\em apriori} any shape
parametrization. By introducing phenomenological shell corrections one
obtains minima of deformation energy for binary fission of parent nuclei at
a finite (non-zero) mass asymmetry. Results are presented for reflection
asymmetric saddle point shapes of thorium and uranium even-mass isotopes
with $A=226$-238 and $A=230$-238 respectively.
\end{abstract}


\maketitle

\section{Introduction}

One of the earliest features observed in nuclear fission was the preference
for breakup into two fragments of unequal mass \cite{hod95mb}. The
asymmetric distribution of fragment masses from the spontaneous or low
excitation energy induced fission was a longstanding puzzle of the theory
\cite{p195b96b}.
In the framework of the liquid drop model (LDM), the mass distribution of
fission fragments is symmetric. By adding shell corrections within
Strutinsky's \cite{str67np} macroscopic-microscopic method, 
it was shown \cite{mol70pl,pas71np} that the
outer barrier for asymmetric shapes is lower than for symmetric ones.
In this
way it was possible to explain qualitatively the fission asymmetry.
Significant progress was achieved with the development of the
asymmetric two center shell model by the Frankfurt school
\cite{mar72zp,gre96bt}.
The fragmentation theory was succesful in describing both regions of low and
high mass asymmetry.
 
The shapes during a fission process from one parent nucleus to the
final fragments, have been intensively studied either statically (looking
for the minimimum of potential energy) \cite{coh63ap,str62jetf},
or dynamically (by choosing a path with the smallest value of action
integral) \cite{hil58c,bra72rmp}.
Particularly important points on a potential energy surface 
are those corresponding to the
ground state~\cite{mun02jnrs}, 
saddle point(s)~\cite{coh63ap,mol00prc} 
and scission point                       
\cite{nag96pl}. 
In a static approach, the equilibrium nuclear shapes are usually
determined by minimizing the energy functional on a certain class of trial
functions representing the surface equation. The required number of
independent shape parameters may be 
as high as nine values \cite{coh63ap}; they are discussed in ref.
\cite{mol00prc}. 

The purpose of this paper is to present a method allowing to obtain a very
general  
reflection asymmetric saddle point shape as a solution of an 
integro-differential equation without a shape parametrization {\em apriori} 
introduced. 
This equation was derived by minimizing the potential energy with constraints 
(constant volume and given deformation parameter). 
The method \cite{str62jetf,p241jnrs02}  allows 
to obtain straightforwardly the axially symmetric surface shape
for which the liquid drop energy, $E_{LDM}=E_s + E_C$,
is minimum. By adding the shell correction $\delta E$ 
to the LDM deformation energy, $E_{def}=E_{LDM} + \delta E$, we can
obtain minima at a finite value of the mass asymmetry parameter.
Phenomenological shell corrections are used. Results for binary fission of
parent nuclei $^{226-238}$Th and $^{230-238}$U are
presented. 

\section{Deformation Energy}
One assumes cylindrical symmetry.
The deformation parameter $\alpha$ is 
defined as the distance between centers of mass
of the fragments lying at the left hand side and right hand side of the
plane $z=0$, respectively: $\alpha=|z^c_L|+|z^c_R|$. 
This definition allows to reach all intermediate stages of deformation from one 
parent nucleus to two fragments by a continuous variation of its value.
The position of separation plane, $z=0$, is given by the condition:
$\left(d\rho/dz\right)_{z=0}=0$.

We are looking for a nuclear surface equation 
$\rho=\rho(z)$  in cylindrical coordinates, which minimizes
the potential energy of deformation with two constraints: volume
conservation, and given deformation parameter, $\alpha$,
assumed to be an adiabatic variable.

Relative to a spherical shape, the deformation energy is defined by
\begin{equation}
E_{def}(\alpha)-E^0= E_s^0[B_s - 1 + 2X(B_C -1)]+\delta E - \delta E^0
\end{equation}
where $E_s^0=a_s(1-\kappa I^2)A^{2/3}$ and $E_C^0 = a_CZ^2A^{-1/3}$ are
energies corresponding to spherical
shape and $I=(N-Z)/A$. The relative surface and Coulomb energies
$B_s=E_s/E_s^0$, $B_C=E_C/E_C^0$ and the shell correction $\delta E(\alpha)$
are functions of the nuclear shape. 
The dependence on the neutron and proton numbers is contained
in $E_s^0$, the fissility parameter $X=E_C^0/(2E_s^0)= [3Z^2e^2/(5R_0)]/2[a_s(1
-\kappa I^2)A^{2/3}]$, as well as in shell correction energy $\delta E^0$
of the spherical nucleus. 
From a fit to experimental data on nuclear masses, quadrupole
moments, and fission barriers, the following values of the parameters
have been obtained~\cite{mye66np}:
$a_s=17.9439$~MeV, $\kappa =1.7826$, $a_C=3e^2/(5r_0)=0.7053$~MeV. 
The radius of spherical nucleus is $R_0=r_0A^{1/3}$ with $r_0=$~1.2249~fm, 
and $e^2=1.44$~MeV$\cdot$fm is the square of electron charge. 
The shape-dependent dimensionless surface term is proportional to the
surface area, and the expression of Coulomb energy is a double integral.
Both are computed~\cite{p75cpc78} by Gauss-Legendre numerical quadratures.

To the LDM energy we add a {\it phenomenological shell correction} $\delta
E(\alpha)$ 
by using a formula adapted after Reference \cite{mye66np}. At a given
deformation we calculate the volumes of fragments and the corresponding
numbers of nucleons $Z_i(\alpha), \ N_i(\alpha)$ ($i=1,2$), proportional 
to the volume of each fragment. Then we add for each fragment the
contribution of protons and neutrons
\begin{equation}
\delta E(\alpha) = \sum_i \delta E_i(\alpha) = \sum_i [\delta E_{pi}(\alpha) 
+ \delta E_{ni}(\alpha)] 
\end{equation}
given by
\begin{equation}
 \delta E_{pi}=Cs(Z_i) ; \ \ \delta E_{ni}=Cs(N_i)
\end{equation}
where
\begin{equation}
 s(Z) = Z^{-2/3}F(Z) -cZ^{1/3} 
\end{equation}
and a similar equation for $s(N)$.
\begin{equation}
F(n) = \frac{3}{5}\left [\frac{N_i^{5/3} -N_{i-1}^{5/3}}{N_i -
N_{i-1}}(n -N_{i-1}) - n^{5/3}+ N_{i-1}^{5/3} \right ]
\end{equation}
where $n \in (N_{i-1}, N_i)$ is the current number of protons ($Z$) or
neutrons ($N$) and $N_{i-1}, N_i$ are the nearest magic numbers. The
parameters $c=0.2$, $C=6.2$~MeV were determined by fit to experimental masses
and deformations.
The dependence on deformation $\alpha$ \cite{sch71pl} is given by
\begin{equation}
\delta E(\alpha) = \frac{C}{2}\left \{ \sum_i[s(N_i)+
s(Z_i)]\frac{L_i(\alpha)}{R_i} \right \}
\end{equation}
where $L_i(\alpha)$ are the lenghths of fragments along the symmetry axis.
During the deformation process, the variation of separation distance between
centers induces the variation of the geometrical quantities and of the
corresponding nucleon numbers. Each time a proton or neutron number reaches
a magic value, the correction energy passes through a minimum, and it has a
maximum at midshell.

\section{Integro-Differential Equation}\label{equ}

As mentioned above, the nuclear surface equation with axial symmetry around
$z$ axis is expressed as $\rho=\rho(z)$ in cylindrical coordinates.
We use the following relationships for the principal radii of curvature
${\cal{R}}_1=\tau \rho$, ${{\cal R}_2}^{-1}=-\rho''/\tau^3$, in which
$\tau^2=1+\rho'^2$.
In order to minimize the deformation energy the surface equation should be a
solution of the following equation
\begin{equation}
\rho\rho''- \rho'^2- \left[ \lambda_1 + \lambda_2 
|z|+ 10XV_s(z,\rho) \right]\rho (1+\rho'^2)^{3/2} - 1 = 0
\end{equation}
where $\rho'=d\rho/dz$, $\rho''=d^2\rho/dz^2$, and
$V_s$ is the Coulomb potential on the nuclear surface.
In this equation $\lambda_1$ and $\lambda_2$ are Lagrange multipliers
corresponding to the constraints of volume conservation (or given mass
asymmetry if the voulume is conserved in each ``half'' of the nucleus) 
and a determined value of the  deformation parameter $\alpha$.
All lengths 
are given in units of $R_0$, Coulomb potential
in units of $Ze/R_0$, and energy in units of the surface energy $E_s^0$.
One can calculate for every value of $\alpha$ the deformation energy
$E_{def}(\alpha)$. The particular value $\alpha_s$ for which 
$dE_{def}(\alpha _s)/d\alpha=0$
corresponds to the extremum,  i.e. the shape function describes the saddle
point (or the ground state). The associated surface equation gives the
unconditional extremum of the energy and corresponds to the fission barrier. 
The other surfaces (for $\alpha \neq \alpha_s$) are extrema only with
condition $\alpha=$~constant.

The Coulomb potential on the surface depends on the function $\rho(z)$, 
hence eq~7 is an integro-differential one, as
$V_s$ is expressed by an integral on the nuclear volume.
The integration method used to solve eq~7 is based on the weak 
dependence of Coulomb energy on the nuclear shape. 
It is invariant under subtraction from $V_s$ of a linear 
function because $\lambda_1$ and $\lambda_2$ are arbitrary constants.
The extremal surface depends on the quantity with which the Coulomb potential
on the nuclear surface differs from the function $\lambda_1 + \lambda_2|z|$, 
where the constants $\lambda_1, \lambda_2$ could be chosen in a way to
minimize this difference. In the next iterration one uses the solution 
$\rho(z)$ previously determined.

The following boundary conditions have to be fulfilled
\begin{equation}
\rho(z_1)=\rho(z_2)=0
\end{equation}
\begin{equation}
\lim_{z \rightarrow z_1}d\rho(z)/dz=\infty \; \; ; \; \;
\lim_{z \rightarrow z_2}d\rho(z)/dz=-\infty
\end{equation}
where $z_1$ and $z_2$ are the intercepts with $z$ axis at the two tips.
Equations~9 are called transversality conditions.
For {\em reflection symmetric shapes} $z_1=-z_2=-z_p$, hence one can consider
only positive values of $z$ in the range $(0,z_p)$.
In order to get rid of singularities in eq~9 it is 
convenient to introduce  a new function
$u(v)$ instead of $\rho(z)$ 
\begin{equation}
u(v )={{\cal{A}}}^2\rho^2(z(v ))
\end{equation}
where
\begin{equation}
z(v )=z_p - v/{\cal{A}}
\end{equation}
By substituting into eq~7 one has
\begin{eqnarray}    
u''-2- & \frac{1}{u}\left[ u'^2+ \left( \frac{5XV_s}{2{\cal{A}}}+
\frac{\lambda_1
+\lambda_2z_p}{4{\cal{A}}} - \frac{\lambda_2v}{4{\cal{A}}^2} \right) 
\right . \cdot \nonumber \\  & \left . (4u+u'^2)^{3/2}\right] =0
\end{eqnarray}    
Then we introduce a linear function of $v$ by adding and subtracting
$a+bv$ to $5XV_s/(2{\cal{A}})$ and defining
$V_{sd}$ as  deviation of Coulomb potential at 
the nuclear surface from a linear function of $v$
\begin{equation}
V_{sd}=[5X/(2{\cal{A}})]V_s -a -vb
\end{equation}
The linear term may be considered an external potential of deformation 
with $a=[5X/(2{\cal{A}})]V_s(v=0)$ and 
$b=\left \{[5X/(2{\cal{A}})]V_s(v=v_p)-a\right\}/v_p$ 
leading to 
\begin{eqnarray}    
u''-2- & \frac{1}{u}\left\{u'^2+ \left[
\left(\frac{\lambda_1+\lambda_2z_p}{4{\cal{A}}}+
a \right) + v \left(b - \frac{\lambda_2}{4{\cal{A}}^2} \right)
\right . \right . \nonumber \\ & \left . \left . 
 + V_{sd} \right](4u+
u'^2)^{3/2}\right\}=0
\end{eqnarray}
Here we have 
new constants ${\cal{A}}$ and $z_p$ 
related to eq~10, besides the previous ones $\lambda_1$ and $\lambda_2$.
Nevertheless the solution is not dependent on each parameter;
important are the linear coefficients in $v$ of the binomial term within
parantheses. By equating with 1 the coefficient of $v$, one can establish 
the following link between parameter ${\cal{A}}$ and the
Lagrange multiplier $\lambda_2$
\begin{equation}
{\cal{A}}^2=\lambda_2/4(b -1)
\end{equation}
In this way $u(v)$ should be determined by equation
\begin{equation}
u'' = 2+\frac{1}{u}[u'^2+(v - d + V_{sd})(4u+u'^2)^{3/2}]
\end{equation}
containing a single parameter $d$.
At the limit 
\begin{equation}
u(0)=0, \; \; u'(0)=1/d
\end{equation}
and eq~9 is satisfied if $z_p=v_p/{\cal{A}}$ is obtained from
\begin{equation}
u '(v _{pn})=0
\end{equation}
The subscript $n$ was introduced as a consequence of the fact that
the number of points $v_{pn}$ (depending on $d$ and other parameters), 
satisfying eq~18 is larger than unity. 
In order to solve eq~16 one starts with given values of parameters
$d$ and $n$. Different classes of shapes solutions of eq.~16 are obtained by
taking various values of $n$: for $n=2$ there is
one neck (binary fission), $n=3$ gives two necks (ternary fission), etc. For
reflection symmetric shapes
$d_L=d_R$ and $n_L=n_R$. Although the parameter ${\cal{A}}$ is not present
in this equation we have to know it in order to determine the shape
function from eq~10. From the volume conservation one has
\begin{equation}
{\cal{A}} = \left\{ \frac{3}{2} \int_0^{v _{pn}} 
u (v )dv \right\}^{1/3}
\end{equation}

After solving the integro-differential 
equation one can find the deformation parameter
$\alpha=z^c_L+z^c_R$, where
\begin{equation}
z^c_L=\int_{z_1}^0|z|\rho^2(z)dz/\int_{z_1}^{z_2}\rho^2(z)dz
\end{equation}
\begin{equation}
z^c_L=\frac{3}{2}{\cal{A}}^{-4}\int_0^{v_{p}}(v_{p} - v)u(v)dv
\end{equation}
depends on $d$. From $\alpha(d)$, one can obtain the inverse function
$d=d(\alpha)$.

For {\em reflection asymmetrical shapes}
we need to introduce another constraint: the asymmetry parameter,
$\eta$, defined by
\begin{equation}
\eta=\frac{M_L-M_R}{M_L+M_R} = \frac{A_1 - A_2}{A_1 + A_2}
\end{equation}
should remain constant during  variation of 
the shape function $\rho(z)$.
Consequently eq~16 should be written differently for left hand
side and right hand side. 
Now $d_L$ is different from $d_R$, and ${\cal{A}}_L \neq {\cal{A}}_R$. 
They have to fulfil matching conditions
$\rho_L(z=0)=\rho_R(z=0)$
hence
\begin{equation}
u_L^{1/2}(v_p)/{\cal{A}}_L = u_R^{1/2}(v_p)/{\cal{A}}_R
\end{equation}
The second derivative $\rho''(z)$  can have a discontinuity in $z=0$
if $d_L \neq d_R$.
The parameters ${\cal{A}}_L$ and ${\cal{A}}_R$ 
are expressed in terms of $\eta$, if we write eq~22 as
\begin{equation}
M_L=\frac{2\pi}{3}(1+\eta)=\pi {\cal{A}}_L^{-3}\int_0^{v_p}u_L(v)dv
\end{equation}
\begin{equation}
M_R=\frac{2\pi}{3}(1-\eta)=\pi {\cal{A}}_R^{-3}\int_0^{v_p}u_R(v)dv
\end{equation}
We assume that $M_L+M_R$ is equal to the mass of a sphere with $R=1$. 
From eqs~24,~25 we obtain
\begin{equation}
{\cal{A}}_L=(1+\eta)^{-1/3}{\cal{A}}_{L0}
\end{equation}
\begin{equation}
{\cal{A}}_R=(1-\eta)^{-1/3}{\cal{A}}_{R0}
\end{equation}
where we introduced notations similar to eq~19
\begin{equation}
{\cal{A}}_{L0(R0)}=\left\{ \frac{3}{2}\int_0^{v_p}u_{L(R)}(v)dv 
\right \}^{1/3}
\end{equation}

The equation 16 is solved by successive approximations. 
In every iteration one uses the 2nd order 
Runge-Kutta numerical method with constant integration step.
The initial value $u''(v=0)$ can be found
straightforwardly by removing the indetermination in
the point $v=0$, 
$u''(0)=-2+(1-b+g)/2d^2$, 
where 
$g=[5X/(2{\cal{A}})][dV_s(v)/dv]_{v=0}$. 
The equation is integrated up to the point $v=v_{pn}$, in which
the first derivative $u'(v_{pn})$ vanishes. 

\section{Mass Asymmetry}

The variations of the saddle point energy with the mass asymmetry parameter
$d_L - d_R$ (which is almost linear function of the mass asymmetry $\eta$)
for some even-mass isotopes of Th and U are plotted in figures~1~and~2. The
minima of the saddle-point energy occur at nonzero mass asymmetry parameters
$d_L-d_R$ between about 0.04 and 0.085 for these nuclei. When the mass
number of an isotope increases, the value of the mass asymmetry
corresponding to the minimum of the SP energy decreases.
\begin{figure}[ht] 
\includegraphics[width=8cm]{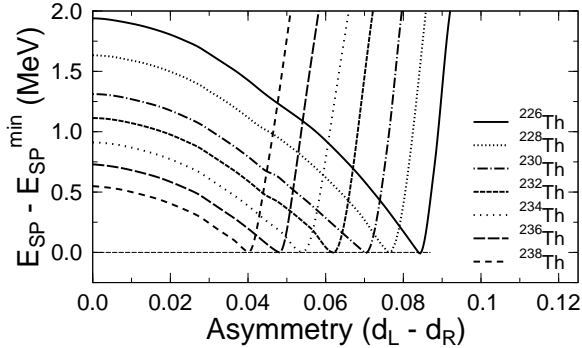} 
\caption{Saddle-point deformation energy versus mass asymmetry parameter for
the binary fission of some even-mass Th isotopes in the presence of shell
corrections.}
\end{figure}
\begin{figure}[hb]
\includegraphics[width=8cm]{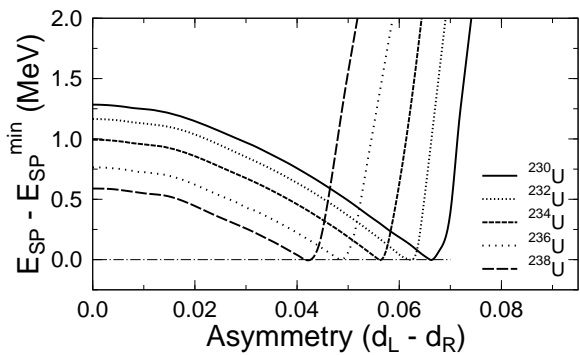} 
\caption{Saddle-point deformation energy versus mass asymmetry parameter for
the binary fission of some even-mass U isotopes in the presence of shell
corrections.}
\end{figure}

From the saddle point energies $E_{SP}$ of every nucleus we subtract the
minimum $E_{SP}^{min}$, as in the figures~1~and~2.  The equilibrium
conditions are changed when the shell effects are taken into account.  The
minimum of the $E_{SP}$ is produced by the negative values of the shell
corrections $\delta E - \delta E^0$.

In conclusion, the liquid drop model saddle point shapes and energy barrier
heights are well reproduced by the present method.
By adding shell corrections to the LDM energy we succeded to
obtain the minima shown in Figures~1~and~2 at a finite value of mass
asymmetry for the binary fission of $^{226-238}$Th and $^{230-238}$U nuclei.

As mentioned by Wilkins et al. \cite{wil76pr}, calculations of PES for
fissioning nuclei ``qualitatively account for an asymmetric division of
mass''. From the qualitative point of view the results displayed in Figures
1 and 2 proove the capability of the method to deal with fission mass and
charge asymmetry. 

The experimentally determined mass number of the most probable heavy
fragment \cite{wah88adnd} for U isotopes ranges from 134 to 140.  The
corresponding values at the displayed minima in Figures 1 and 2 are very
close to 125, which means a discrepancy between 6.7~\% and 10.7~\% for
$A_H$. 
The inaccuracy in reproducing the experimental mass asymmetry is due to the
contribution of the phenomenological shell corrections. In the absence of
shell corrections the pure liquid drop model (LDM) reflection-symmetric
saddle point shapes [8] are reproduced, and the barrier height increseas
with an increased mass asymmetry. When the shell corrections are taken into
account the LDM part behaves in the same manner (larger values at non-zero
mass asymmetry). Only the contribution of shell effects can produce a
minimum of the barrier height at a finite value of the mass asymmetry.
One may hope to obtain a better agreement with experimental data by
using a more realistic shell correction model.

\begin{acknowledgments}
This work was partly supported by UNESCO (UVE-ROSTE Contract 875.737.2),
by the Centre of
Excellence IDRANAP under contract ICA1-CT-2000-70023 with European
Commission, Brussels, by
Bundesministerium f\"{u}r Bildung und Forschung (BMBF), Bonn, Gesellschaft
f\"{u}r Schwerionenforschung (GSI), Darmstadt,
by Ministry of Education and Research, Bucharest.
\end{acknowledgments}


\begin{thebibliography}{33}

\bibitem{hod95mb}  
D. C. Hoffman, T. M. Hamilton and M. R. Lane, in {\em Nuclear Decay Modes},
(Institute of Physics, Bristol, 1996) p. 393.

\bibitem{p195b96b}
D. N. Poenaru and W. Greiner (Eds), {\em Nuclear Decay Modes}, (Institute of
Physics, Bristol, 1996). 

\bibitem{str67np}
V. M. Strutinsky, Nucl. Phys., {\bf A95}, 420 (1967).

\bibitem{mol70pl}
P. M{\"o}ller and S. G. Nilsson, Phys. Lett., B{ \bf 31}, 283 (1970).

\bibitem{pas71np}
V. V. Pashkevich, Nucl. Phys., { \bf A 169}, 275 (1971).

\bibitem{mar72zp}
J. A. Maruhn and W. Greiner,  Z. Phys.{ \bf 251}, 431 (1972); 
Phys. Rev. Lett.{ \bf 32}, 548 (1974).

\bibitem{gre96bt} W. Greiner and J. A. Maruhn, {\em Nuclear Models}
(Springer, Berlin, 1996).

\bibitem{coh63ap} S. Cohen and W. J. Swiatecki, Ann. Physics (N. Y.) 
{\bf 22}, 406 (1963).

\bibitem{str62jetf}
V. M. Strutinsky, JETF {\bf 42}, 1571 (1962).

\bibitem{hil58c}
D. L. Hill,  {\em Proc. of the 2nd U. N. Int. Conf. on the Peaceful
Uses of Atomic Energy} (United Nations, Geneva, 1958), p~244.

\bibitem{bra72rmp}
M. Brack, J. Damgaard, A. Jensen, A., H. C. Pauli, V.~M.~Strutinsky
and G.~Y.~Wong,  Rev. Mod. Phys.{ \bf 44}, 320 (1972).

\bibitem{mun02jnrs}
I. Muntian, Z. Patyk, A. Sobiczewski, J. Nucl. Radiochem. Sci. (Japan),
{\bf 3},  169 (2002).

\bibitem{mol00prc} P. M\"oller and A. Iwamoto, Phys. Rev. C {\bf 61}, 
047602 (2000); P. M\"oller, D. G. Madland, A. J. Sierk, and A. Iwamoto,
Nature {\bf 409}, 785 (2001). 

\bibitem{nag96pl}
Y. Nagame et al.,  Phys. Lett. {\bf 387}, 26 (1996); 
I. Nishinaka et al.,
Phys. Rev. C {\bf 56}, 891 (1997); 
Y. L. Zhao et al., 
Phys. Rev. C {\bf 62}, 014612 (2000);  
Phys. Rev. Lett. {\bf 82}, 3408 (1999).

\bibitem{p241jnrs02}
D.~N. Poenaru, W. Greiner, Y. Nagame, and R.~A. Gherghescu,
J. Nucl. Radiochem. Sci. (Japan), {\bf 3}, 43 (2002).

\bibitem{mye66np}
W.~D. Myers and W.~J. Swiatecki, Nucl. Phys.  {\bf A81}, 1  (1966).

\bibitem{p75cpc78}
D.~N. Poenaru and M. Iva\c{s}cu,  Comp. Phys. Communic. {\bf 16}, 85
(1978); 
D.~N. Poenaru, M. Iva\c{s}cu, and D. Mazilu,  Comp. Phys. Communic.
{\bf 19}, 205  (1980).

\bibitem{sch71pl}
H. Schultheis and R. Schultheis, Phys. Lett. B {\bf 37}, 467 (1971).

\bibitem{wil76pr} P. D. Wilkins, E. P. Steinberg and R. R. Chasman, 
Phys. Rev. C {\bf 14}, 1832 (1976).

\bibitem{wah88adnd} A. C. Wahl,  Atomic Data Nucl. Data Tables 
{\bf 39}, 1 (1988).

\end{thebibliography}
\end{document}